# Optimal dense coding and swap operation between two coupled electronic spins: effects of nuclear field and spin-orbit interaction

Li Jiang(姜丽)[1], Guo-Feng Zhang(张国锋)[1,2,3,*]

[1]*Key Laboratory of Micro-Nano Measurement-Manipulation and Physics (Ministry of Education), School of Physics and Nuclear Energy Engineering; State Key Laboratory of Software Development Environment, Beihang University, Xueyuan Road No. 37, Beijing 100191, China*

[2]*State Key Laboratory of Low-Dimensional Quantum Physics, Tsinghua University, Beijing 100084, China*

[3]*Key Laboratory of Quantum Information, University of Science and Technology of China, Chinese Academy of Sciences, Hefei 230026, China*

**Abstract:** The effects of nuclear field and spin-orbit interaction on dense coding and swap operation are studied in detail for both the antiferromagnetic (AFM) and ferromagnetic (FM) coupling cases. The conditions for a valid dense coding and under which swap operation is feasible are given.

**PACS number(s):**   03.67. Lx; 03.65.Ud; 75.10.Jm

**Keywords:** Optimal dense coding; swap operation; two coupled electronic spins

## I. Introduction

Entanglement has been extensively studied in recent years because of the intriguing features of quantum mechanics, and plays a fundamental role in quantum information processing, such as quantum key distribution [1], quantum teleportation [2], dense coding [3] and so on. According to the original dense coding scheme [4], the sender can transmit two bits of classical information to the receiver by sending a single qubit if they share a two-qubit maximally entangled state (an Einstein-Podolsky-Rosen (EPR) state). There are many schemes to realize qubit, however, the spins of electrons or nuclei are the natural candidates to represent qubits since they are natural binary systems.

Many researches on dense coding have been conducted experimentally [5] or theoretically [6-8]. In an ordinary dense coding, the sender performs one of the local unitary transformations $U_i \in U(d)$ on $d$-dimensional quantum system to put the initially shared

---

entangled state $\rho$ in $\rho_i = (U_i \otimes I_d)\rho(U_i^+ \otimes I_d)$ with a priori probability $p_i (i = 0,1,...i_{max})$, and then the sender sends off his quantum state to the receiver. Upon receiving this quantum system, the receiver performs a suitable measurement on $\rho_i$ to extract the signal. Holevo quantity [9], which can be described as $\chi = S(\bar{\rho}) - \sum_{i=0}^{i_{max}} p_i S(\rho_i)$, is an ideal pattern to bind the optimal amount of information that can be conveyed. Here $S(\rho)$ denotes the von Neumann entropy and $S(\bar{\rho}) = \sum_{i=0}^{i_{max}} p_i \rho_i$ is the average density matrix of the signal ensemble. We can use Holevo quantity as the definition of the capacity of dense coding since it is asymptotically achievable [10]. Further, the von Neumann entropy is invariable under unitary transformations, i.e. $S(\rho_i) = S(\rho)$. Hence, the dense coding capacity can be rewritten as $\chi = S(\bar{\rho}) - S(\rho)$. The next step is to find the optimal signal ensemble $\{\rho_i; p_i\}_{i=0}^{i_{max}}$ that maximizes $\chi$. In Ref. [11], the author showed that the $d^2$ signal states $(i_{max} = d^2 - 1)$ generated by mutually orthogonal unitary transformations with equal probabilities yield the maximum, which is called optimal dense coding, and considered the optimal dense coding when the shared entangled state was a general mixed one. In this paper, we will take the so called thermal entangled state [12] as the shared one between the sender and receiver to investigate the optimal dense coding.

Another fundamental condition required for quantum computation (QC) is the universal quantum gates that implement the unitary transformations [13]. Of the various schemes that have been proposed, the ones based on solid state systems are believed to have the best scalability. Moreover, the solid state schemes can largely take advantage of modern semiconductor technology and micro-fabrication technology [14]. The minimal requirements for a quantum computer architecture are the existence of fundamental quantum bits and the ability to carry out qubit operations, such as the quantum exclusive or gate [also known as controlled-not (CNOT)], the Walsh–Hadamard gate and the swap gate, which is defined by $U_{swap}|\Phi\rangle \otimes |\Psi\rangle = |\Psi\rangle \otimes |\Phi\rangle$ [15]. The swap operation is a particularly intriguing process and is the most non-local operation and can act as a double-teleportation [16], although it transforms product states to product states. The square root of a swap gate $\sqrt{U_{swap}}$ is universal while the swap gate itself is not universal. A CNOT gate can be achieved through a combination of single-qubit operations and $\sqrt{U_{swap}}$ [17]:

$$U_{CNOT} = e^{i\frac{\pi}{4}\sigma_1^z} e^{-i\frac{\pi}{4}\sigma_2^z} U_{swap}^{1/2} e^{i\frac{\pi}{2}\sigma_1^z} U_{swap}^{1/2}. \tag{1}$$

Since spin itself can be used as qubits not only in some real physical systems but also in many other systems, such as a superconductor, quantum dots, and a trapped ions, we can use two coupled electronic spins system to investigate the effects of nuclear field and spin-orbit interaction on dense coding and the qualification for implementing the swap gate. In the proposed spin-based QC architectures, the exchange interaction between spins plays a fundamental role in the establishment of two-qubit thermal entangled states which will be shared by the sender and receiver, while the Zeeman splitting, which is a function of the external magnetic field, provides various single-qubit operations.

The paper is organized as follows. In Sec. II we present the model and its analytical solution. The optimal dense coding based on the thermal state associated with this model is investigated in Sec. III. Sec. IV is dedicated to considering the conditions that the swap operation is feasible. Our final conclusion remarks are presented in Sec. V.

**II. The model and solutions**

Now we consider two coupled electronic spins in the presence of nuclear field and spin-orbit interaction

$$H = H_{exc} + H_{hf}$$
$$= J\left[S_1 \cdot S_2 + \beta_0 \cdot (S_1 \times S_2)\right] + \gamma_e\left[(B_{ext} + B_{1,n}) \cdot S_1 + (B_{ext} + B_{2,n}) \cdot S_2\right], \tag{2}$$

where $J$ is the exchange constant ($J>0$ corresponds to AFM case and $J<0$ FM case), $S_1 S_2$ is the isotropic exchange interaction, $\beta_0 (S_1 \times S_2)$ is the anisotropic spin-orbit interaction which comes from Dzyaloshinkii-Moriya [18] (DM) interaction and $\beta_0$ is the $z$-component of the DM vector coupling, $B_{ext}$ is the external magnetic field and $B_n$ is the effective nuclear magnetic field, the subscripts 1 and 2 refer to the spin 1 and 2, respectively. By defining $B=B_{ext}+ (B_{1,n}+B_{2,n})/2$ and $dB= (B_{1,n}-B_{2,n})/2$, Eq. (2) can be written as

$$H = J\left[S_1 \cdot S_2 + \beta_0 \cdot (S_1 \times S_2)_z\right] + \gamma_e\left[B \cdot (S_1 + S_2) + dB \cdot (S_1 - S_2)\right]. \tag{3}$$

In the computational basis $\{|\uparrow,\uparrow\rangle, |\uparrow,\downarrow\rangle, |\downarrow,\uparrow\rangle, |\downarrow,\downarrow\rangle\}$ the Hamiltonian becomes

$$H = \frac{J}{4}\begin{pmatrix} 1 & 0 & 0 & 0 \\ 0 & -1 & 2(1+i\beta_0) & 0 \\ 0 & 2(1-i\beta_0) & -1 & 0 \\ 0 & 0 & 0 & 1 \end{pmatrix} + \gamma_e \begin{pmatrix} 2B_z & \Theta^*+d\Theta^* & \Theta^*-d\Theta^* & 0 \\ \Theta+d\Theta & -2dB_z & 0 & \Theta^*-d\Theta^* \\ \Theta-d\Theta & 0 & 2dB_z & \Theta^*+d\Theta^* \\ 0 & \Theta-d\Theta & \Theta+d\Theta & -2B_z \end{pmatrix}, \quad (4)$$

where, $\Theta=B_x+iB_y$, $d\Theta=dB_x+idB_y$ and * stand for complex conjugate. In the following, we consider the common approximation of a large external magnetic field along the $z$ direction so $B_{ext,z}\gg B_n$, where the common values used in experiments and theoretical works are $B_{ext,z}\approx 100$ and $B_n\approx 1$ to 5mT, which implies a large energy gap between $\Theta$, $d\Theta$ and $B_z$, $dB_z$ which makes the transition probabilities between $|\uparrow,\uparrow\rangle \to \genfrac{|}{\rangle}{0pt}{}{\uparrow,\downarrow}{\downarrow,\uparrow}$ and $|\downarrow,\downarrow\rangle \to \genfrac{|}{\rangle}{0pt}{}{\uparrow,\downarrow}{\downarrow,\uparrow}$ very small, that allows to reduce the Hamiltonian to

$$H = \begin{pmatrix} \frac{J}{4} - \gamma_e B_z & 0 & 0 & 0 \\ 0 & -\frac{J}{4} - \gamma_e dB_z & \frac{J}{2}(1+i\beta_0) & 0 \\ 0 & \frac{J}{2}(1-i\beta_0) & -\frac{J}{4} + \gamma_e dB_z & 0 \\ 0 & 0 & 0 & \frac{J}{4} + \gamma_e B_z \end{pmatrix}. \quad (5)$$

For simplicity, we define $|1\rangle$ and $|0\rangle$ as the spin-up and spin-down states, respectively. The eigenstates and corresponding eigenvalues of the Hamiltonian are given by

$$|\psi_1\rangle = |11\rangle, E_1 = \frac{J}{4} - \gamma_e B_z;$$

$$|\psi_4\rangle = |00\rangle, E_4 = \frac{J}{4} + \gamma_e B_z;$$

$$|\psi_2\rangle = \frac{\eta_- i}{J(i+\beta_0)\sqrt{\xi_+}}|10\rangle + \frac{1}{\sqrt{\xi_+}}|01\rangle, E_2 = -\frac{J}{4} - \frac{\sqrt{J_{eff}}}{2};$$

$$|\psi_3\rangle = \frac{\eta_+ i}{J(i+\beta_0)\sqrt{\xi_-}}|10\rangle + \frac{1}{\sqrt{\xi_-}}|01\rangle, E_3 = -\frac{J}{4} + \frac{\sqrt{J_{eff}}}{2}, \quad (6)$$

with $\eta_\pm = -2\gamma_e dB_z \pm \sqrt{J_{eff}} = -dB_{zeff} \pm \sqrt{J_{eff}}$, $\xi_\pm = 1 + \left(\sqrt{J_{eff}} \pm dB_{zeff}\right)^2 / J^2(1+\beta_0^2)$ and

$J_{eff} = J^2(1+\beta_0^2) + dB_{zeff}^2$. As thermal fluctuation is introduced into the system, the so-called thermal state at equilibrium (temperature $T$) is $\rho = \frac{1}{Z}\sum_l e^{-\beta E_l}|\psi_l\rangle\langle\psi_l|$, $Z = 2e^{-J/4T}\cosh[\gamma_e B_z/T] + 2e^{J/4T}\cosh[\sqrt{J_{eff}}/2T]$ is the partition function, $\beta = 1/(k_B T)$ and $k_B$ is the Boltzmann constant. Here we write $k_B = 1$ for simplicity.

### III. Optimal dense coding

We conduct the optimal dense coding with the thermal entangled states of two coupled electronic spins system as a channel. The set of mutually orthogonal unitary transformations [19] of the optimal dense coding for two-qubit is

$$U_{0,0}|x\rangle = |x\rangle; U_{1,0}|x\rangle = e^{\sqrt{-1}(\pi/2)x}|x\rangle;$$
$$U_{0,1}|x\rangle = |x+1(\bmod 2)\rangle; U_{1,1}|x\rangle = e^{\sqrt{-1}(\pi/2)x}|x+1(\bmod 2)\rangle, \quad (7)$$

where $|x\rangle$ is the single qubit computational basis $(|x\rangle = |0\rangle, |1\rangle)$.

The average state of the ensemble of signal states generated by the unitary transformations Eq. (7) is

$$\bar{\rho} = \frac{1}{4}\sum_{i=0}^{3}(U_i \otimes I_2)\rho(U_i^+ \otimes I_2). \quad (8)$$

We have assumed 0→00; 1→01; 2→10; 3→11, and $\rho$ is the thermal state associated with Eq. (6). Through straightforward algebra, we have

$$\bar{\rho} = \frac{1}{4}\big[|00\rangle\langle 00| + |01\rangle\langle 01| + |10\rangle\langle 10| + |11\rangle\langle 11|\big]. \quad (9)$$

After completing the set of mutually orthogonal unitary transformations, the maximum dense coding capacity $\chi$ can be written as

$$\chi = S(\bar{\rho}) - S(\rho) = 2 - S(\rho), \quad (10)$$

where $S(\rho)$ is the von Neumann entropy of the quantum state $\rho$. Thus, the value of the maximal dense coding capacity becomes

$$\chi = \frac{Z\ln[4] + A + B - Z\ln[Z]}{Z\ln[2]}, \quad (11)$$

with $A = e^{J/4T} \left( J \cosh[\sqrt{J_{eff}}/2T] + 2\sqrt{J_{eff}} \sinh[J_{eff}/2T] \right)/2T$, and

$B = e^{-J/4T} \left( -J \cosh[\gamma_e B_z/T] + 4\gamma_e B_z \sinh[\gamma_e B_z/T] \right)/2T$. In order to investigate the influence of nuclear field, we set $B_z=0$, which implies $B_{ext}=0$ and $B_{1,n}+B_{2,n}=0$ according to Eq. (3). Thus, $dB_z = (B_{1,n} - B_{2,n})_z/2$. Under the above simplified condition, the Eq. (11) can be written as

$$\chi = \frac{2J + 4T \ln[4] - 4T \ln[2\zeta] + 2\delta/\zeta}{4T \ln[2]}, \quad (12)$$

where $\zeta = 1 + e^{J/2T} \cosh[\sqrt{J_{eff}}/2T]$ and $\delta = -J + e^{J/2T} \sqrt{J_{eff}} \sinh[\sqrt{J_{eff}}/2T]$. It is easily found that optimal dense coding capacity $\chi$ satisfy $\chi(dB_{zeff})=\chi(-dB_{zeff})$. From executed effectively point of view, in order to conduct the optimal dense coding successfully, the parameters of the model must satisfy

$$\chi > \log_2 2 = 1 \Leftrightarrow e^{\frac{J}{2T}} \left( J \cosh[\frac{\sqrt{J_{eff}}}{2T}] + \sqrt{J_{eff}} \sinh[\frac{\sqrt{J_{eff}}}{2T}] \right) > 2\zeta T \ln[\zeta]. \quad (13)$$

The effects of $J$, $\beta_0$, $dB_{zeff}$ and $T$ on $\chi$ will be analyzed in the following.

When the temperature is zero, the thermal state becomes $|\Phi\rangle_\pm = \pm(\sqrt{J_{eff}} + dB_{zeff})\lambda(\beta_0 - i)/J(1+\beta_0^2)|10\rangle \pm \lambda i|01\rangle$ ($\lambda = \sqrt{1/2(1 - dB_{zeff}/\sqrt{J_{eff}})}$) which is a pure state. So the maximum dense coding capacity $\chi$ will be **2** since $S(\rho)$ is zero for a pure state. However, by increasing the temperature, the thermal state will be mixed and dense coding capacity becomes smaller. In Fig.1, the optimal dense coding capacity $\chi$ as a function of the coupling constant $J$ and effective nuclear magnetic field $dB_{zeff}(=2\gamma_e dB_z)$ is plotted for a definite temperature and DM vector coupling $\beta_0$. We find optimal dense coding capacity $\chi$ is symmetric with respect to nuclear magnetic field $dB_{zeff}$, which can be easily understood since $\chi$ depends only on the quadratic term of $dB_{zeff}$. Moreover, the FM coupling is more suitable than AFM coupling for a valid dense coding and nuclear magnetic field only has a weaker effect for FM case than for AFM one.

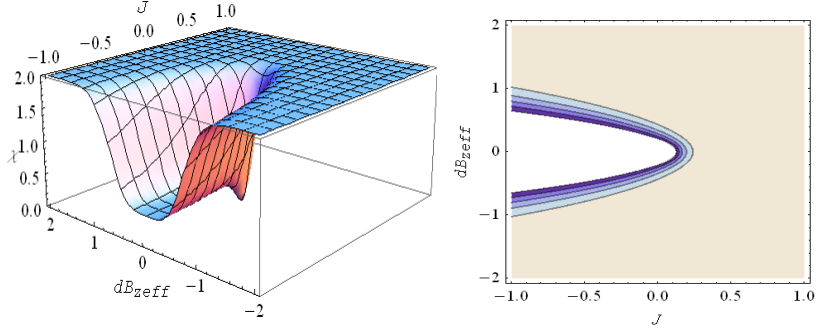

Fig.1: Optimal dense coding capacity $\chi$ versus coupling constant $J$ and effective nuclear magnetic field $dB_{zeff}$, we assume $T=0.05$ and $\beta_0=0.01$, the right panel is the contour.

In order to investigate the influence of nuclear magnetic field on the dense coding for AFM case, $\chi$ is given as a function of $dB_{zeff}$ for different DM coupling strength $\beta_0$ in Fig.2. On the other hand, if the spin-orbit coupling is turned off, the thermal state can be described as

$$\rho = \begin{pmatrix} \dfrac{1}{Z_0} & 0 & 0 & 0 \\ 0 & \dfrac{e^{J/2T}(\sqrt{J_{eff}}+dB_{zeff}\tanh[\theta])}{2\sqrt{J_{eff}}(e^{J/2T}+\sec h[\theta])} & -\dfrac{Je^{J/2T}\sinh[\theta]}{Z_0\sqrt{J_{eff}}} & 0 \\ 0 & -\dfrac{Je^{J/2T}\sinh[\theta]}{Z_0\sqrt{J_{eff}}} & \dfrac{e^{J/2T}(\cosh[\theta]-dB_{zeff}\sinh[\theta]/\sqrt{J_{eff}})}{Z_0} & 0 \\ 0 & 0 & 0 & \dfrac{1}{Z_0} \end{pmatrix},$$

where $Z_0 = 2+2e^{J/2T}\cosh[\theta]$ and $\theta = \sqrt{J_{eff}}/2T$. Obviously, the thermal state is mixed; in general, the dense coding capacity cannot arrive at 2. From Fig.2, it is seen that optimal dense coding capacity $\chi$ becomes larger until arrives at **2** with the increasing of DM coupling. Further, from Eq.(12), we know DM coupling has the same influence as nuclear magnetic field $dB_{zeff}$ on $\chi$ since $\beta_0$ and $dB_{zeff}$ have the same dependence on $\chi$ for $J=\pm 1$. This result can be seen from Fig.3, where $J=-1$ and is the same with the right part of Fig.2. With the increasing of DM coupling strength, for example, when $\beta_0$ is infinite, the thermal state becomes $|\Phi\rangle = \sqrt{2}/2(|10\rangle + i|01\rangle)$ for an AFM coupling and $|\Phi\rangle = \sqrt{2}/2(|10\rangle - i|01\rangle)$ for a FM coupling, they are maximally entangled pure states. So $\chi$ will arrive at 2.

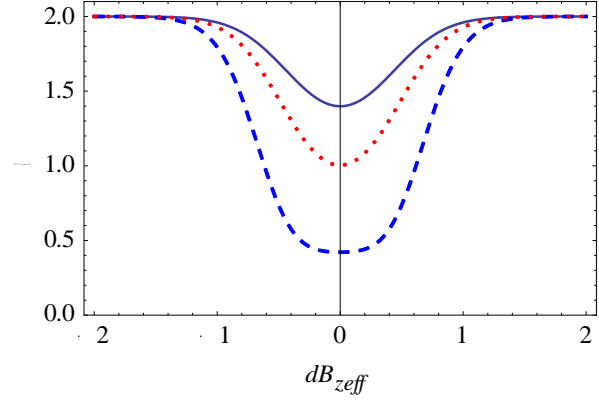

Fig.2: Optimal dense coding capacity $\chi$ versus the effective nuclear magnetic field $dB_{zeff}$ for FM (J=-1) case. From top to bottom, the DM interaction $\beta_0$ is 0.8, 0.65, and 0.2, respectively, $T$=0.05.

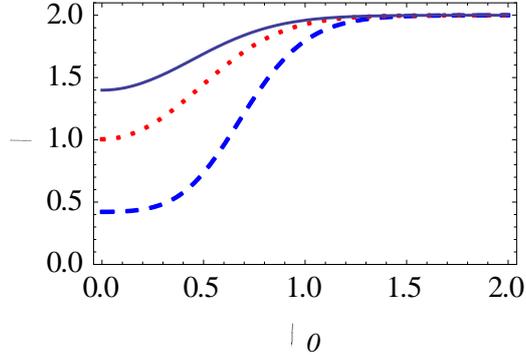

Fig.3: Optimal dense coding capacity $\chi$ versus the DM vector coupling $\beta_0$ for FM (J=-1). From top to bottom, the effective nuclear magnetic field $dB_{zeff}$ is 0.8, 0.65 and 0.2 respectively, $T$=0.05.

The relation between optimal dense coding capacity $\chi$ and temperature $T$ are described in Fig.4. It is obvious to understand that $\chi$ has more opportunities to be ideal value 2 with the lower temperature both for AM and AFM case. Moreover, we find $\chi$ will decrease quickly with the increasing of the temperature for the FM coupling; while falls for AFM case.

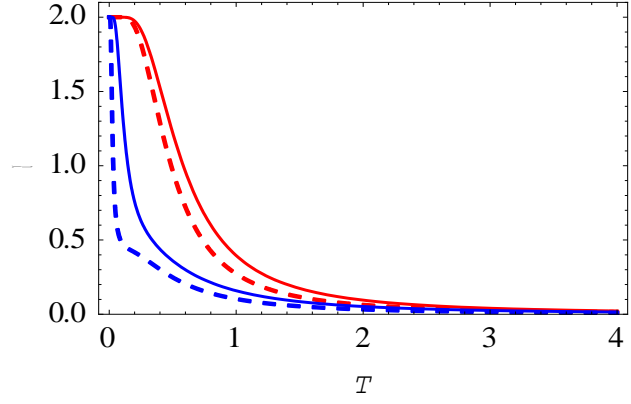

FIG.4: Optimal dense coding capacity $\chi$ versus the temperature $T$. The left two lines correspond to a FM case ($J$=-1) and the right correspond to an AFM case ($J$=1). From top to bottom, nuclear magnetic field $dB_{zeff}$ is 1.2, 0.5, respectively.

**VI. Swap operation**

In order to investigate the swap operation, the initial state is chosen as a product state given by

$$\psi(0)=(\alpha_1|1\rangle+\beta_1|0\rangle)\otimes(\alpha_2|1\rangle+\beta_2|0\rangle)=\begin{pmatrix}\alpha_1\\ \beta_1\end{pmatrix}\otimes\begin{pmatrix}\alpha_2\\ \beta_2\end{pmatrix}. \tag{14}$$

And it then evolves under the Hamiltonian (5):

$$\psi(t)=e^{-iHt}\psi(0). \tag{15}$$

If the wavefunction becomes $(\alpha_2|1\rangle+\beta_2|0\rangle)\otimes(\alpha_1|1\rangle+\beta_1|0\rangle)$ at some time, then the swap operation is achieved. From the above expressions, swap is achieved by exchanging the coefficients of the unpolarized state $|01\rangle$ and state $|10\rangle$. Expand Eq. (5) in the basis $\{|11\rangle,|10\rangle,|01\rangle,|00\rangle\}$

$$\psi(t)=a[t]|11\rangle+b[t]|10\rangle+c[t]|01\rangle+d[t]|00\rangle. \tag{16}$$

With

$$a[t]=\alpha_1\alpha_2 e^{-iE_1 t}; b[t]=e^{iJt/4}\left(P_+ e^{it\sqrt{J_{eff}}/2}+P_- e^{-it\sqrt{J_{eff}}/2}\right)/2;$$

$$d[t]=\beta_1\beta_2 e^{-E_4 t}; c[t]=e^{iJt/4}\left(Q_- e^{it\sqrt{J_{eff}}/2}+Q_+ e^{-it\sqrt{J_{eff}}/2}\right)/2.$$

And

$$P_{\pm} = \left(1 \pm \frac{2dB_{zeff}\gamma_e}{\sqrt{J_{eff}}}\right)\alpha_1\beta_2 \mp \frac{J(1+i\beta_0)}{\sqrt{J_{eff}}}\beta_1\alpha_2; Q_{\pm} = \left(1 \pm \frac{2dB_{zeff}\gamma_e}{\sqrt{J_{eff}}}\right)\beta_1\alpha_2 \pm \frac{J(1-i\beta_0)}{\sqrt{J_{eff}}}\alpha_1\beta_2 .$$

If a two-qubit system is in a disentangled state, the reduced density matrix of either spin is pure. The reduced density matrix of the first spin is given by:

$$\rho_{1,11} = |a[t]|^2 + |b[t]|^2; \rho_{1,00} = |c[t]|^2 + |d[t]|^2;$$
$$\rho_{1,10} = a[t]c[t]^* + b[t]d[t]^*; \rho_{1,01} = a[t]^*c[t] + b[t]^*d[t], \quad (17)$$

where * stand for complex conjugate. The eigenvalue equation for $\rho_1$ is

$$\tau^2 - (\rho_{1,11} + \rho_{1,00})\tau + (\rho_{1,11}\rho_{1,00} - |\rho_{1,10}|^2) = 0 . \quad (18)$$

To achieve a swap operation, we must have a product state of spin 1 and 2 evolve into a product state, and the Schmidt number of the two-spin state cannot exceed one, which means that only one eigenvalue of the reduced density $\rho_1$ is non-vanishing, so $\rho_{1,11}\rho_{1,00} - |\rho_{10}|^2 = 0$. From Eq. (18) we have:

$$\rho_{1,11}\rho_{1,00} - |\rho_{1,10}|^2 = |a[t]d[t] - b[t]c[t]|^2 = -e^{-iJt/2}(\mu + iv)/2J_{eff} , \quad (19)$$

where

$$\mu = X\alpha_2^2\beta_1^2 + 2Y\alpha_1\alpha_2\beta_1\beta_2;$$
$$X = e^{iJt}J(-i+\beta_0)\left(2i\gamma_e dB_z(-1+\cos[\sqrt{J_{eff}}t]) + \sqrt{J_{eff}}\sin[\sqrt{J_{eff}}t]\right);$$
$$Y = -J^2(1+\beta_0^2) + 4dB_z^2(-1+e^{iJt})\gamma_e^2 + e^{iJt}J^2(1+\beta_0^2)\cos[\sqrt{J_{eff}}t];$$
$$v = e^{iJt}J(i+\beta_0)\left(2i\gamma_e dB_z(-1+\cos[\sqrt{J_{eff}}t]) + i\sqrt{J_{eff}}\sin[\sqrt{J_{eff}}t]\right)\alpha_1^2\beta_2^2.$$

Now we consider the value of $t$ that makes Eq. (19) vanish. If $t$ depends on the initial state parameters $\alpha_1, \alpha_2, \beta_1$ and $\beta_2$, then for an unknown initial state the swap operation cannot be realized. Hence, $t$ must be independent of the initial state parameters. The conditions that make Eq. (19) vanish can be divided into two cases:

Case 1: $\sqrt{J^2(1+\beta_0^2) + dB_{zeff}^2}\, t = 2k\pi$ $(k=0,1,2...)$ and $Jt = 2n\pi$ $(n=0,\pm 1,\pm 2...)$.

Case 1.1: when $k = n$ the solution is only $t = 0$, thus $|\psi_1\rangle = \alpha_1|1\rangle + \beta_1|0\rangle$.

Case 1.2: for $k > n$ then $t = 2\pi\sqrt{(k^2 - n^2)/(J^2(1+\beta_0^2) + dB_{zeff}^2)}$. If $k+n$ is even, $|\psi_1\rangle = \alpha_1|1\rangle + \beta_1|0\rangle$. However, if $k = n$ is odd, $|\psi_1\rangle = \alpha_1|1\rangle + e^{-i\pi}\beta_1|0\rangle$, the state of first spin returns to the initial one except for an additional phase shift $e^{-i\pi}$. When $k < n$, there is no solution. Hence, swap operation cannot be achieved in Case 1.

Case 2:

$$\sqrt{J^2(1+\beta_0^2) + dB_{zeff}^2}\, t = (2k+1)\pi \ (k = 0,1,2...) \text{ and } Jt = (2n+1)\pi \ (n = 0, \pm 1, \pm 2...)$$

and $dB_{zeff} = 0$.

Case 2.1: for $k = n$, the solution is only $t = 0$, $|\psi_1\rangle = \alpha_1|1\rangle + \beta_1|0\rangle$ the state of first spin is the initial one.

Case 2.2: when $k > n$, $t = \sqrt{(k+n+1)(k+n)}\,2\pi/J\beta_0$. If $k+n$ is even,

$$|\psi_1\rangle = \alpha_2|1\rangle + e^{i\arccos[\frac{2n+1}{2k+1}]}\beta_2|0\rangle \text{ for } J > 0 \text{ and } |\psi_1\rangle = \alpha_2|1\rangle + e^{i(\pi+\arccos[\frac{2n+1}{2k+1}])}\beta_2|0\rangle \text{ for } J < 0.$$

So, we can see that the states of the two spins are swapped except for an additional phase shift (different for AFM case and FM case), so that the swap operation is achieved after the additional phase shift is corrected by a single-spin operation. If $k+n$ is odd,

$$|\psi_1\rangle = \alpha_2|1\rangle + e^{i(\pi+\arccos[\frac{2n+1}{2k+1}])}\beta_2|0\rangle \text{ for } J > 0 \text{ and } |\psi_1\rangle = \alpha_2|1\rangle + e^{i\arccos[\frac{2n+1}{2k+1}]}\beta_2|0\rangle \text{ for } J < 0.$$

As achievement of a swap operation, the scenario is similar as the case of $k+n$ is even except the additional phase shift is exchanged for AFM and FM case. However, for $k < n$, there is no solution.

**V. Conclusions**

To summarize, we have investigated the optimal dense coding and swap operation in a coupled electronic spins model. The effect of temperature $T$, spin-orbit coupling $\beta_0$ and effective nuclear magnetic field $dB_{zeff}$ on optimal dense coding capacity $\chi$ for both AFM and FM cases are studied in detail. We found $\chi$ is symmetric with respect to $dB_{zeff}$ whenever AFM or FM. Moreover, the FM coupling is more suitable than AFM coupling for a dense coding and nuclear magnetic field only has a weaker effect for FM case than for AFM one. Spin-orbit coupling $β_0$ and effective nuclear magnetic field $dB_{zeff}$ have the same influence pattern on χ. As for swap operation achievement base on this model, there are numerous strict conditions to be followed. The conditions that must be needed for a swap operation are given in great detail.

**Acknowledgments**

This work is supported by the National Natural Science Foundation of China (Grant No. 11574022 and 11174024) and the Open Fund of IPOC (BUPT) grants Nos. IPOC2013B007, also supported by the Open Project Program of State Key Laboratory of Low-Dimensional Quantum Physics (Tsinghua University) grants Nos. KF201407, and Beijing Higher Education (Young Elite Teacher Project) YETP 1141.